\documentclass[conf]{new-aiaa}
\usepackage[utf8]{inputenc}

\usepackage{subcaption}
\usepackage{graphicx}
\usepackage{amsmath}
\usepackage{makecell}

\usepackage{float}

\usepackage{tikz}
\usetikzlibrary{circuits.ee.IEC}
\usetikzlibrary{shapes,arrows,calc,positioning}
\tikzstyle{bigblock} = [draw, fill=blue!20, rectangle, 
    minimum height=6em, minimum width=8em]
\tikzstyle{medblock} = [draw, fill=blue!20, rectangle, 
    minimum height=4em, minimum width=4em]    
\tikzstyle{mux} = [draw, fill=black!20, rectangle, 
    minimum height=5em, minimum width=0.1em]    
\tikzstyle{smallblock} = [draw, fill=blue!20, rectangle, 
    minimum height=2em, minimum width=3em]
    
\tikzstyle{data_block} = [draw, fill=green!20, rectangle, 
    minimum height=2em, minimum width=3em]
\tikzstyle{ops_block} = [draw, fill=blue!20, rectangle, 
    minimum height=2em, minimum width=3em]    
\tikzstyle{est_block} = [draw, fill=red!20, rectangle, 
    minimum height=2em, minimum width=3em]    
    
\tikzstyle{sum} = [draw, fill=blue!20, circle, node distance=1cm,minimum height=0.5cm]
\tikzstyle{signal} = [coordinate]
\tikzstyle{pinstyle} = [pin edge={to-,thin,black}]
\tikzstyle{block} = [draw, fill=blue!20, rectangle, 
    minimum height=3em, minimum width=9em]
\tikzstyle{blockS} = [draw, fill=blue!20, rectangle, 
    minimum height=3em, minimum width=4em]    
\tikzstyle{input} = [coordinate]
\tikzstyle{output} = [coordinate]

\tikzstyle{gain} = [draw, fill=blue!20, regular polygon, regular polygon sides=3, shape border rotate=30]
\tikzstyle{gain_vert} = [draw, fill=blue!20, regular polygon, regular polygon sides=3, shape border rotate=0]

\usetikzlibrary{matrix}

\usetikzlibrary{positioning}
\usetikzlibrary{math}

\usepackage{hyperref}
\usepackage{xcolor}
\hypersetup{
    colorlinks,
    linkcolor={blue!100!black},
    citecolor={blue!50!black},
    urlcolor={blue!80!black}
}

\newcommand{\bc}{\begin{center}}
\newcommand{\ec}{\end{center}}
\newcommand{\benum}{\begin{enumerate}}
\newcommand{\eenum}{\end{enumerate}}
\newcommand{\nn}{\nonumber}
\newcommand{\matl}{\left[ \begin{array}}
\newcommand{\matr}{\end{array} \right]}

\renewcommand{\matl}{\begin{bmatrix}}
\renewcommand{\matr}{\end{bmatrix}}

\newcommand{\matls}{\left[ \begin{smallmatrix}}
\newcommand{\matrs}{\end{smallmatrix} \right]}
\newcommand{\isdef}{\stackrel{\triangle}{=}}

\newcommand{\rmE}{{\rm E}}

\newcommand{\rmP}{{\rm P}}

\newcommand{\rmc}{{\rm c}}

\newcommand{\rme}{{\rm e}}

\newcommand{\rmp}{{\rm p}}

\newcommand{\BBR}{{\mathbb R}}

\newcommand{\BBZ}{{\mathbb Z}}

\newcommand{\SW}{{\mathcal W}}

\usepackage{enumitem}
\newlist{todolist}{itemize}{2}
\setlist[todolist]{label=$\square$}
\usepackage{pifont}

\usepackage[colorinlistoftodos,textwidth=1 in,textsize=footnotesize]{todonotes}
\title{Feedback Linearization-based Guidance Law \\ for Guaranteed Interception}

\author{
Alexander Dorsey%
\footnote{Undergraduate Research Assistant, Department of Mechanical Engineering, University of Maryland, Baltimore County, 1000 Hilltop Circle, Baltimore, MD 21250.} and
Ankit Goel\footnote{Assistant Professor, Department of Mechanical Engineering, University of Maryland, Baltimore County, 1000 Hilltop Circle, Baltimore, MD 21250.}
}

\begin{document}
\maketitle

\begin{abstract}
    This paper presents an input-output feedback linearization (IOL)-based guidance law to ensure interception in a pursuer-evader engagement scenario. 
    A point-mass dynamic model for both the pursuer and the evader is considered.
    An IOL guidance law is derived using range and line-of-sight (LOS) rate measurements.
    It is found that the range-based IOL guidance law exhibits a singularity under certain conditions. 
    To address this issue, a fuzzy logic system is employed to smoothly blend the IOL guidance with the classical proportional guidance law, thereby avoiding the singularity.
    In contrast, the LOS-based IOL guidance law is free of singularities but suffers from divergence issues due to angle-related complications.
    To resolve this, a simple correction function is introduced to ensure consistent interception behavior.
    Results from Monte Carlo simulations indicate that both modifications of the IOL guidance laws cause interception with control limits applied. 
\end{abstract}

\section{Introduction}

Missile guidance remains a challenging problem due to nonlinear dynamics, nonminimum phase behavior resulting from nose-mounted gyro sensors and tail-fin actuation, actuator constraints, and uncertain aerodynamic loading. 
The proportional guidance law is the most widely used guidance system \cite{zarchan2012tactical} due to its mathematical simplicity and low computational cost. 
However, proportional guidance law is derived under strict assumptions in the kinematic engagement scenario, which are easily violated in real-world scenarios. 

Optimal control frameworks have been investigated to develop optimal guidance laws for guaranteed interception \cite{chen2010optimal}. 
However, due to the computational complexity and offline nature of the solution techniques, optimal strategies are unsuitable for real-time implementation. 
Recently, polynomial guidance laws which depend on the estimate of the time-to-go metric have been investigated \cite{tahk2019augmented}.
However, these approaches do not yield stable closed-loop dynamics and thus cannot guarantee interception. 

This paper explores the application of the input-output feedback linearization (IOL) framework to develop guidance laws that can guarantee interception \cite{isidori1985nonlinear}. 
The IOL technique is briefly described in \cite{portella2024circumventing, delgado2024adaptive}.
The IOL framework was investigated to develop guidance laws for launch vehicles in \cite{burchett2005feedback} and for a kinematic pursuer-evader scenario in \cite{alkaher2014guidance}. 
In this paper, the IOL framework is used to derive guidance law for the interception problem modeled with point-mass dynamics for both the pursuer and the evader.
In particular, two IOL-based guidance systems are developed using range and line-of-sight (LOS) measurements, respectively.
However, both approaches face implementation challenges. 
The range-based guidance system exhibits a singularity in the nominal tail-chase scenario.
In contrast, the LOS-based guidance system can produce trajectories that diverge from the evader under certain conditions. 
The main contribution of this paper is the development of modified IOL-based guidance laws that overcome these limitations for both range- and LOS-based formulations.
% 
% A fuzzy system blends the range-based IOL guidance law with the classical proportional guidance law to yield a guidance system that does not exhibit a singularity. 
To address the singularity in the range-based IOL guidance law, a fuzzy system is used to blend it with the classical proportional guidance law, resulting in a robust hybrid guidance strategy free from singularities. 
% A LOS-based correction system is proposed for the LOS-based guidance system that ensures that all trajectories of the pursuer converge towards the evader irrespective of the initial configuration.  
For the LOS-based guidance system, a correction function is proposed that guarantees convergence of the pursuer's trajectory to the evader, regardless of the initial engagement geometry.

This paper is organized as follows. 
Section \ref{interception_dynamics} reviews the planar interception dynamics, formulates the IOL guidance laws, and presents the modifications to circumvent the implementation challenges of the baseline IOL guidance laws.
% Section \ref{range_iol} presents the fuzzified range‐based IOL law. Section \ref{los_iol} describes the angle correction algorithm for line‐of‐sight linearization.
Section \ref{simulations} presents numerical simulation results demonstrating the application of the modified IOL guidance laws to the interception problem. 
Finally, the paper concludes with a summary of the preliminary results and plans for the additional work to be reported in the final manuscript. 

\section{Interception Dynamics and Linearizing Guidance}\label{interception_dynamics}

This section briefly reviews the planar interception dynamics and develops the input-output linearizing guidance law to ensure interception.
A detailed derivation of the interception dynamics can be found in \cite{kabamba2014fundamentals}, and a brief description of the input-output feedback linearization technique can be found in \cite{portella2024circumventing}. 

\subsection{Planer Interception Dynamics}
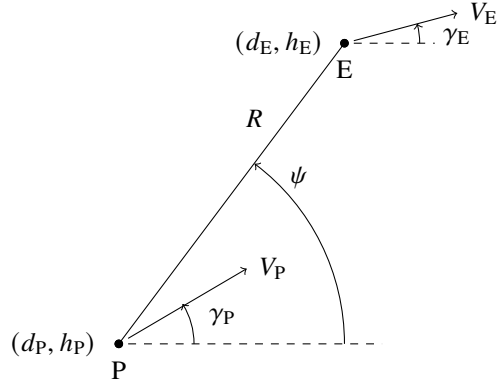
\begin{figure}[h]
% {l}{0.4\textwidth}
    \centering    
    {
    \begin{tikzpicture}
    % [auto, node distance=2cm,>=latex',text centered]

    \node at (0,0) (P) {};    
    \draw [fill=black] (P) circle [radius=0.050];

    \node at (3,4) (E) {};    
    \draw [fill=black] (E) circle [radius=0.050];

    \draw[->] (P.center) 
            node[xshift=0, yshift=-10] {$\rmP$} circle [radius=0.050]
            -- 
            (E.center) node[xshift=0, yshift=-10] {$\rmE$} circle [radius=0.050];

    \draw[->] (P) -- +(1.7, 1) node[xshift=10, yshift=0] {$V_\rmP$};
    \draw[->] (E) -- +(1.5, 0.4) node[xshift=10, yshift=0] {$V_\rmE$};

    \draw[-, dashed] (P) node[xshift=-25, yshift=0] {$(d_\rmP,h_\rmP )$} -- +(3.5, 0);
    \draw[-, dashed] (E) node[xshift=-25, yshift=0] {$(d_\rmE,h_\rmE )$} -- +(1.2, 0);

    \draw [->] (1,0) arc [radius=1, start angle=0, end angle= 32] node[xshift=15, yshift=-5] {$\gamma_\rmP$};

    \draw [->] (4,4) arc [radius=1, start angle=0, end angle= 15] node[xshift=15, yshift=-5] {$\gamma_\rmE$};
    
    \draw [->] (3,0) arc [radius=3, start angle=0, end angle= 53] node[xshift=17, yshift=-5] {$\psi$};

    \node at (1.8,3) (R) {$R$};  
    
    \end{tikzpicture}
    }
    \caption{ Interception geometry for the pursuer and the evader.    
    }
    % \vspace{-8em}
    \label{fig:interception_geometry}
\end{figure}

A planar interception geometry for a pursuer $\rmP$ and an evader $\rmE$ is shown in Figure \ref{fig:interception_geometry} .
The interception dynamics for the pursuer and the evader is 
\begin{align}
    \dot V_\rmP 
        &=
            \dfrac{T_\rmP-D_\rmP}{m_\rmP} - g \sin \gamma_\rmP, \\
    \dot V_\rmE 
        &=
            \dfrac{T_\rmE-D_\rmE}{m_\rmE} - g \sin \gamma_\rmE, \\
    % \dot \gamma_\rmP 
    %     &=
    %         -\dfrac{1}{V_\rmP} (n_{z,\rmP} + g \cos \gamma_\rmP), \\
    \dot \gamma_\rmP 
        &=
            -\dfrac{1}{V_\rmP} \left( n_{z,\rmP} + g \cos \gamma_\rmP\right), 
    \label{eq:dot_gamma_P}        
    \\
    \dot \gamma_\rmE
        &=
            -\dfrac{1}{V_\rmE} \left(n_{z,\rmE} + g \cos \gamma_\rmE \right),
\end{align}
where $V_\rmP, V_\rmE$ are the velocities, 
$\gamma_\rmP,$ $\gamma_\rmE$ are the flight-path angles, 
$T_\rmP, T_\rmE$ are the thrust, 
$D_\rmP, D_\rmE$ are the drag, and 
$n_{z,\rmP}, n_{z,\rmE}$ are the normal acceleration of the pursuer and evader, respectively.
The line of sight angle $\psi$ and the range $R$ satisfy
\begin{align}
    \dot \psi
        &=
            \dfrac{1}{R}
            \left( 
                V_\rmP \sin(\psi - \gamma_\rmP)
                -
                V_\rmE \sin(\psi - \gamma_\rmE)
            \right),
    \label{eq:dotbeta}
    \\
    \dot R 
        &=
            V_\rme  \cos(\psi - \gamma_\rme)
            -
            V_\rmp \cos(\psi - \gamma_\rmp) 
            .
    \label{eq:dotR}
\end{align}
In this work, the drag on the pursuer and the evader is modeled as described in \cite{islam2022minimum}.

\subsection{State-space Reformulation}

To formulate the input-output feedback linearizing guidance law, the equations of motion need to be reformulated in the affine state-space form.  
Defining
\begin{align}
x = \begin{bmatrix}
    R\\ \psi\\ V_\rmP \\ \gamma_\rmP
\end{bmatrix}, \quad 
w = \begin{bmatrix}
    V_\rmE\\
    \gamma_\rmE
\end{bmatrix}, \quad
u = n_{z,\rmP}, 
\end{align}
the governing equations can be written as
\begin{align}
    \dot x = f(x, w) + g(x) u, 
    \label{eq:SS_form}
\end{align}
where 
\begin{align}
    f(x,w) 
        \isdef
    \begin{bmatrix}
        V_\rmE \cos(\psi-\gamma_\rmE) - V_\rmP \cos(\psi - \gamma_P) \\
        -\left(V_\rmE \sin(\psi - \gamma_\rmE) - V_\rmP \sin(\psi - \gamma_\rmP) \right) \\
        \dfrac{T_\rmP - D_\rmP}{m_\rmP} - g \sin(\gamma_\rmP) \\
        \dfrac{-g \cos(\gamma_\rmP)}{V_\rmP} 
    \end{bmatrix}, 
    \quad 
    g(x)
        \isdef
            \begin{bmatrix}
                0\\ 0\\ 0\\ -\dfrac{1}{V_\rmP}
            \end{bmatrix}.
\end{align}
Furthermore, 
\begin{align}
    \dot w = \SW(w),
\end{align}
where
\begin{align}
    \SW(w)
        \isdef 
            \begin{bmatrix}
                \dfrac{T_\rmE - D_\rmE}{m_\rmP} - g \sin(\gamma_\rmE) \\
                -\dfrac{n_{z,E}+g \cos(\gamma_\rmE)}{V_\rmE} 
            \end{bmatrix}.
\end{align}
Note that the state $w$ contains the information of the evader.

\subsection{Range-based Feedback Linearization}\label{range_iol}
First, we consider the case where the output is the range measurement, that is, 
\begin{align}
    y = h(x) = R. 
\end{align}
Note that the relative degree of the range $R$ with respect to the input $n_{z, \rmP}$ is $2$.
Therefore, as shown in \cite{delgado2023circumventingunstablezerodynamics}, the linearized dynamics is
\begin{align}
    \dot \xi &= A_\rmc \xi + B_\rmc v, \\
    y &= C_\rmc \xi, 
\end{align}
where
\begin{align}
    A_\rmc
        &\isdef
            \begin{bmatrix}
                0 & 1\\
                0 & 0
            \end{bmatrix}, 
    \quad
    B_\rmc 
        \isdef
            \begin{bmatrix}
                0\\
                1
            \end{bmatrix}, 
    \quad 
    C_\rmc 
        \isdef 
            \begin{bmatrix}
                1 &
                0
            \end{bmatrix}.
\end{align}
% \todo{Alex, list out the matrices above. }
and the linearizing input $u$ is given by
\begin{align}
    u(x,w)
        =
            \beta(x)^{-1} 
            \left(
                -\alpha(x,w) + v    
            \right),
            \label{eqn:IOL_control_R}
\end{align}
% The functions $\alpha(x,w)$ and $\beta(x)$ are given by
where
\begin{align}
    \alpha(x,w) 
        &\isdef
            L_f^2 h(x)
        =
            \dfrac{\left(V_\rmE \sin(\psi-\gamma_\rmE) - V_\rmP \sin(\psi - \gamma_\rmP)\right)^2}{R} +
            \cos(\psi - \gamma_\rmP)\left(\dfrac{D_\rmP - T_\rmP }{m_\rmP} \right) + g\cos(\gamma_\rmP) \sin(\psi-\gamma_\rmP), 
    \\
    \beta(x)
        &\isdef
            L_g L_f h(x)
        =
            \sin(\psi - \gamma_\rmP).
            \label{eqn:Beta(x)_R}
\end{align}

The internal control signal $v$ is chosen as 
\begin{align}
    v = -k_R \xi_1 = -k_R R,
\end{align}
where $k_R>0,$ which yields the closed-loop dynamics
\begin{align}
    \ddot R + k_R R = 0.
\end{align}
Note that the closed-loop dynamics are not asymptotically stable, but only Lyapunov stable. 
However, since the interception objective only requires the range $R$ to cross zero once, the proposed guidance law is sufficient to guarantee successful interception.
% Note that a simple proportional guidance law is chosen since it will drive $R$ to zero

Note that $\beta(x)$ is not invertible when $\psi - \gamma_\rmP = n \pi,$ where $n \in \BBZ,$ which corresponds to the tail chase or direct head-on course.  
However, since the guidance law \eqref{eqn:IOL_control_R} requires the inversion of $\beta(x), $ the guidance law cannot be practically used since the command signal, which is $n_{z, \rmP}$ in this case, will approach $\pm \infty$ as $\psi - \gamma_\rmP$ approaches $n \pi.$

\begin{figure}[h]
    \centering
    \includegraphics[width=0.55\columnwidth]{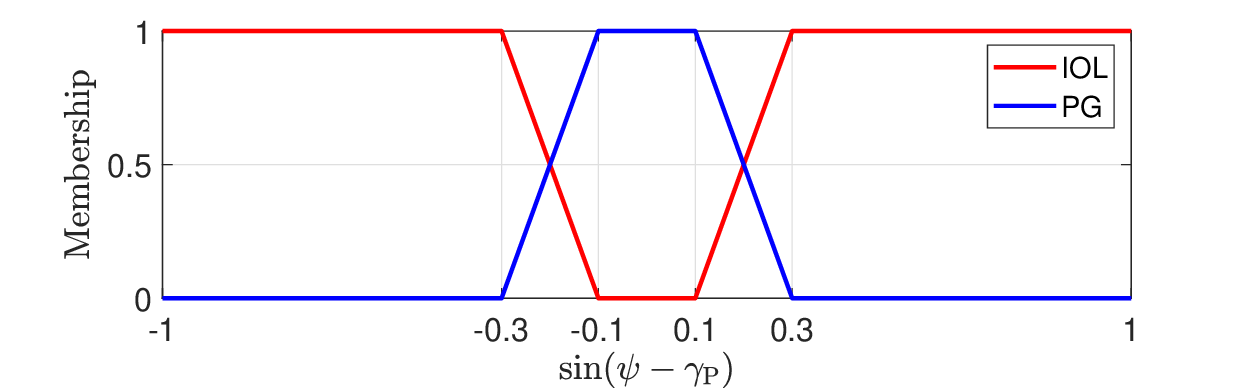}
    \caption{Membership function used in the fuzzy system. The red trace is the membership of the IOL guidance law, and the blue line is the membership of the proportional guidance.}
    \label{fig:FuzzyMemebershipFunctionFigure}
\end{figure}

To circumvent this singularity, we use a fuzzy system to transition the guidance law to the classical proportional guidance in the case when $\psi - \gamma_\rmP$ approaches $n \pi.$
In particular, we use the Takagi Sugeno (T-S) fuzzy system to design a fuzzy logic system which switches between the classical proportional guidance and the IOL guidance law \eqref{eqn:IOL_control_R} as shown in \cite{salazar2024mpc}.
The T-S fuzzy system uses a membership function, shown in Figure \ref{fig:FuzzyMemebershipFunctionFigure}, to combine the output of the IOL guidance law \eqref{eqn:IOL_control_R} and the  classical proportional guidance law \cite{kabamba2014fundamentals} given by
\begin{align}
    u(x,w) 
        =
            -\lambda \frac{V_\rmP}{R} 
            \Big(
                V_\rmP \sin(\psi - \gamma_\rmP) - V_\rmE \sin(\psi - \gamma_\rmE) 
            \Big)
            - g \cos(\gamma_\rmP),
    \label{eqn:PG}
\end{align}
where $\lambda>0$ is the proportional guidance constant, to generate the normal acceleration command.   
Specifically, the output of the fuzzy system is 
\begin{align}
    u(x,w)
        &=
            \mu_{\rm{IOL}}(\sin(\psi - \gamma_\rmP)) \left( \beta(x)^{-1} 
            \left(
                -\alpha(x,w) + v    
            \right) \right)
            \nn \\
            &\quad +
            \mu_{\rm{PG}}(\sin(\psi - \gamma_\rmP))
            \left( -\lambda \frac{V_\rmP}{R} \left(V_\rmP \sin(\psi - \gamma_\rmP) - V_\rmE \sin(\psi - \gamma_\rmE) \right) - g \cos(\gamma_\rmP) \right), 
\end{align}
where 
$\mu_{\rm IOL}, \mu_{\rm PG} : \BBR \to (0,1)$ are membership functions.
Note that, for all $\theta \in \BBR,$ $\mu_{\rm{IOL}}(\theta) + \mu_{\rm{PG}}(\theta)  = 1.$
Since $\mu_{\rm{IOL}}(\sin(\psi - \gamma_\rmP)) = 0$ when $\sin(\psi - \gamma_\rmP)\in (-0.1, 0.1)$ the singularity in \eqref{eqn:IOL_control_R} is circumvented.  

\subsection{Line-of-sight-based Feedback Linearization}\label{los_iol}
Next, we consider the case where the output is the rate of change of line of sight, that is, 
\begin{align}
    y = h(x) = \dot \psi. 
\end{align}
Note that the relative degree in this case is $1$.
Therefore, as shown in \cite{delgado2023circumventingunstablezerodynamics}, the linearized dynamics is
\begin{align}
    \dot \xi &= A_\rmc \xi + B_\rmc v, \\
    y &= C_\rmc \xi, 
\end{align}
where
$
    A_\rmc
        \isdef
            0, 
    \
    B_\rmc 
        \isdef
            1, 
$ and $ 
    C_\rmc 
        \isdef 
            1.
$
% \todo{Alex, list out the matrices above. }
and the linearizing input $u$ is given by
\begin{align}
    u(x,w)
        =
            \beta(x)^{-1} 
            \left(
                -\alpha(x,w) + v    
            \right),
            \label{eqn:IOL_control_dotPsi}
\end{align}
where 
\begin{align}
    \alpha(x,w) &= L_f h(x)
    = 
        \dfrac{2\left( V_\rmE \cos(\psi - \gamma_\rmE) - V_\rmP \cos(\psi - \gamma_\rmP)\right) \left(V_\rmE \sin(\psi - \gamma_\rmE) - V_\rmP \sin(\psi - \gamma_\rmP) \right)}{R^2} \notag \\
      &\quad
        - \dfrac{\sin(\psi - \gamma_\rmP) \left(\dfrac{D_\rmP - T_\rmP}{m_\rmP} + g \sin(\gamma_\rmP) \right)}{R} + \dfrac{g\cos(\gamma_\rmP) \cos(\psi - \gamma_\rmP)}{R}, 
    \\
    \beta (x) &= L_g h(x) 
    = \dfrac{\cos(\psi - \gamma_\rmP)}{R}.
\end{align}

The internal control signal $v$ is chosen as 
\begin{align}
    v = -k_{\dot \psi} \xi = -k_{\dot psi} \dot \psi,
\end{align}
where $k_{\dot \psi}>0,$ which yields the asymptotically stable closed-loop dynamics for $\dot \psi,$ that is, 
\begin{align}
    \ddot \psi + k_{\dot \psi} \dot \psi = 0.
\end{align}
Note that the proposed guidance law ensures that $\dot \psi \to 0,$ but does not directly regulate $\psi.$
As shown in Figure \ref{fig:PreLogicFigure}, it is observed that the proposed guidance law, in some scenarios, may yield a pursuer trajectory that ensures $\dot \psi$ converges to zero but $R \to \infty.$
To avoid this scenario, the following correction logic is included in the loop.
\begin{equation}
    u(x,w)
        = \begin{cases}
            -\beta(x)^{-1} (-\alpha(x,w) + v) \quad &,  \ |\gamma_{\rmP} - \psi| > 90, \\
            \beta(x)^{-1} (-\alpha(x,w) + v) \quad &, \ |\gamma_{\rmP} - \psi| < 90.
        \end{cases}
    \label{eq:IOL_control_dotPsi_mod}
\end{equation}

\section{Numerical Results}\label{simulations}

This section presents preliminary results demonstrating the application of the feedback linearization-based guidance law to the interception problem. 
In this work, we assume that the evader has a constant thrust of $50$ $\rm kN$ and a mass of $10000$ $\rm kg$, a drag coeffecent of $0.025$, a frontal surface area of $28$ meters squared, and its only normal acceleration is due to gravity.
Furthermore, at the start of the interception, that is, at $t=0,$ $V_\rmE = 584$, $\rm m /s^{2}$ $\gamma_\rmE = 10$ degrees, $h_\rmE = 10000 \ \rm m$, $ X_\rmE = 5000 \ \rm{m}$.
The pursuer has a constant thrust of $15$ $\rm kN$ and a mass of $204$ $\rm kg$, frontal surface area of $2.3$ $\rm m^2$, and at the start of the interception, $V_\rmP = 800 $ $\rm m/s^2$, $\gamma_\rmP = 1$ $\rm degrees$, $h_\rmP = 5000 \ \rm m$, $ X = 0 \ \rm{m}$.
As seen in the lower-right subplot, the fuzzy system selects the appropriate guidance law as required to guarantee interception. 

\subsection{Comparision with Proportional Guidance}
\begin{figure}[h]
    \centering
    \includegraphics[width=0.5\columnwidth]{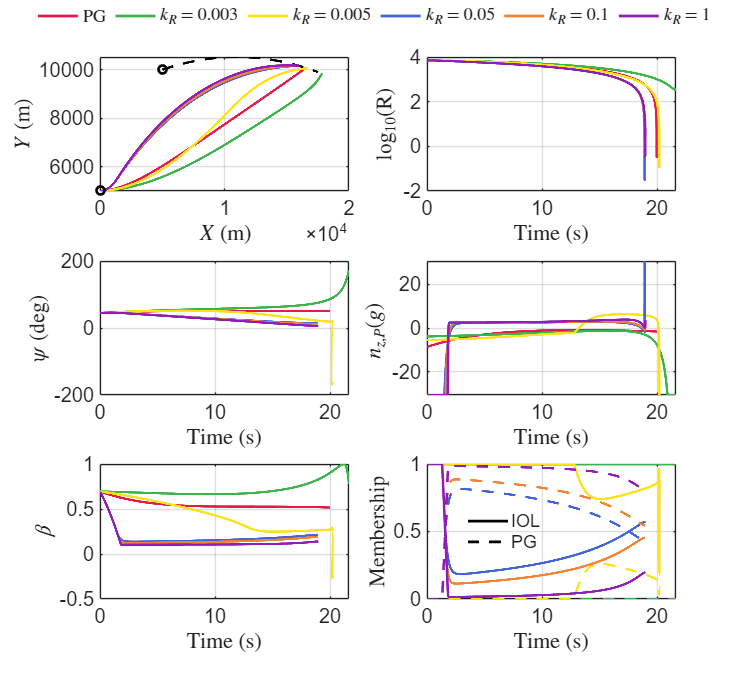}
    \caption{Interception response with range-based IOL guidance law \eqref{eqn:IOL_control_R} and various values of the proportional gain $k_R.$ Proportional Guidance is shown in red for comparison.}
    \label{fig:KpTuningFigure}
\end{figure}

First, we use the range-based feedback linearization guidance law \eqref{eqn:IOL_control_R} to intercept the evader. 
Figure \ref{fig:KpTuningFigure} shows the interception response with various values of the proportional gain $k_R.$
Note that, as expected, the interception is successful for all values of the proportional gain $k_R.$

\begin{figure*}[h!]
    \centering
    \begin{subfigure}[b]{0.5\textwidth}
        \centering
        \includegraphics[width=\columnwidth]{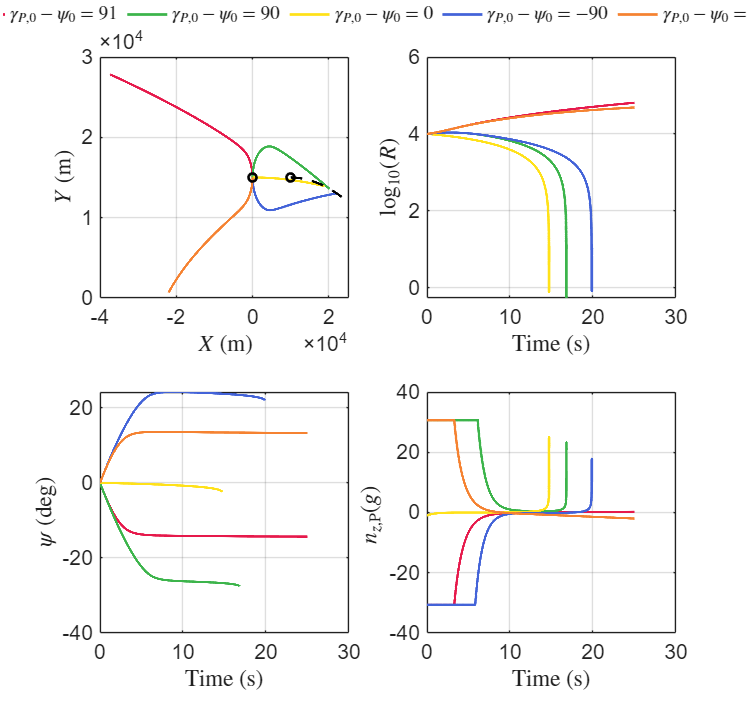}
        \caption{Guidance law \eqref{eqn:IOL_control_dotPsi} without the correction.}
        \label{fig:PreLogicFigure}
    \end{subfigure}%
    % ~ 
    \begin{subfigure}[b]{0.5\textwidth}
        \centering
        \includegraphics[width=\columnwidth]{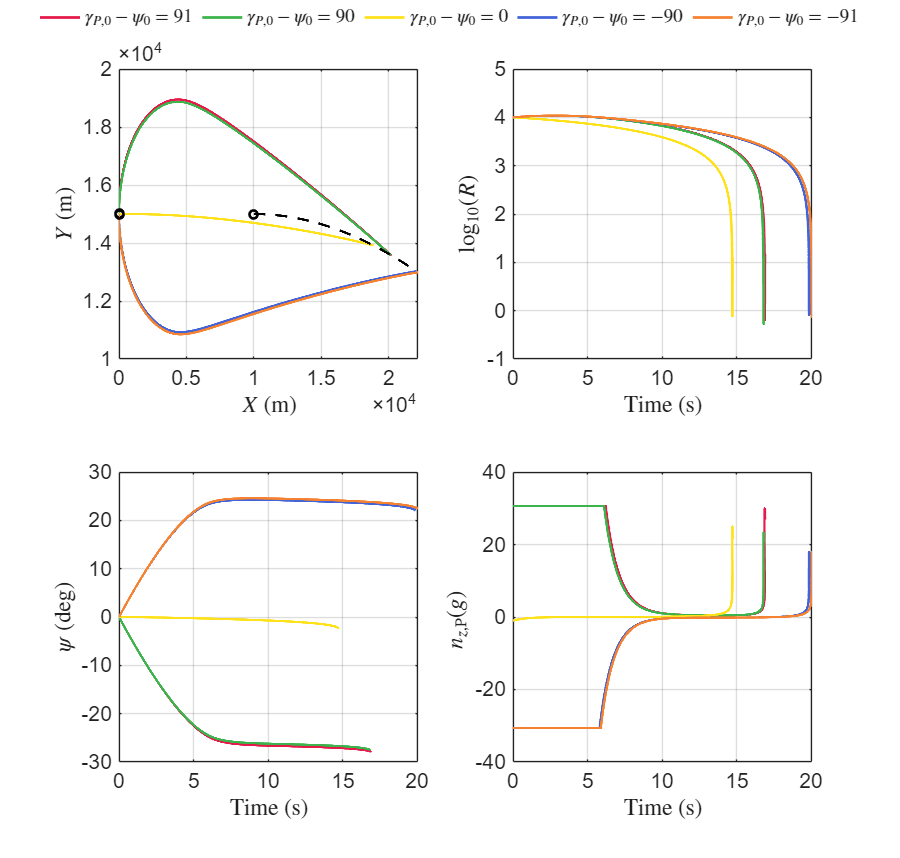}
        \caption{Guidance law \eqref{eq:IOL_control_dotPsi_mod} with the correction.}
        \label{fig:PostLogicFigure}
    \end{subfigure}
    \caption{Interception response with line-of-sight-based IOL guidance law \eqref{eqn:IOL_control_dotPsi} and \eqref{eq:IOL_control_dotPsi_mod}.}
\end{figure*}

Next, we use the line-of-sight-based feedback linearization guidance law to intercept the evader. 
First, we demonstrate the necessity of the correction modification \eqref{eq:IOL_control_dotPsi_mod} to the guidance law \eqref{eqn:IOL_control_dotPsi}.
We consider the case where the initial value of  $\gamma_{\rmP}-\psi,$ that is, $\gamma_{\rmP,0}-\psi_0 \in \{91, 90, 0, -90, -91\}. $ 
Figure \ref{fig:PreLogicFigure} and \ref{fig:PostLogicFigure} show the interception response with line-of-sight-based IOL guidance law \eqref{eqn:IOL_control_dotPsi} and \eqref{eq:IOL_control_dotPsi_mod}, respectively. 
Note that, without the correction, that is, with the guidance law \eqref{eqn:IOL_control_dotPsi}, $R \to \infty$ in the scenario when $|\gamma_{\rmP} - \psi| > 90.$
However, with the proposed correction, that is, with the guidance law \eqref{eq:IOL_control_dotPsi_mod}, $R \to 0$ irrespective of the value of $\gamma_{\rmP} - \psi.$

\begin{figure}[h!]
    \centering
    \includegraphics[width=0.5\columnwidth]{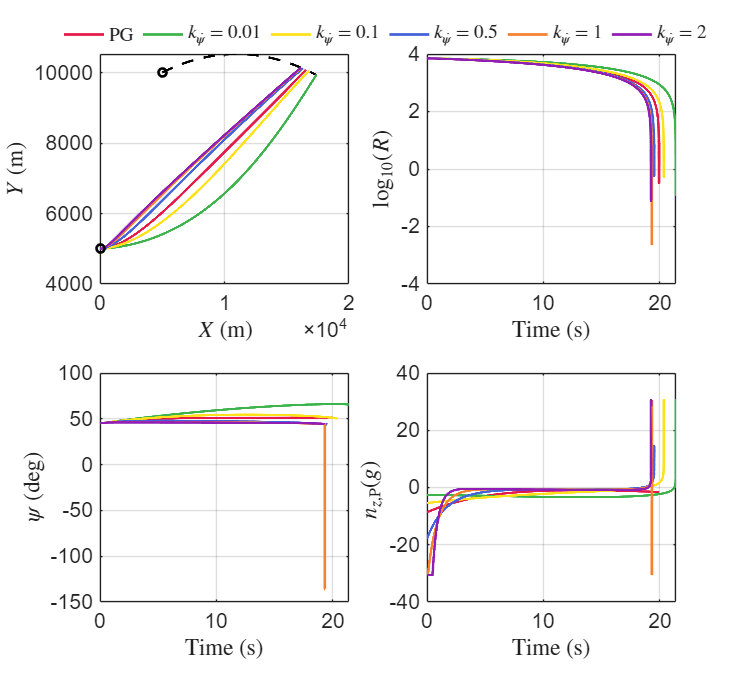}
    % \caption{Interception data throughout various $k_{\dot{\psi}}$ gains. Proportional Guidance (PG) is shown in red.}
    \caption{Interception response with LOS-based IOL guidance law \eqref{eqn:IOL_control_dotPsi} and various values of the proportional gain $k_{\dot \psi}.$ Proportional Guidance is shown in red for comparison.}
    \label{fig:LQRTuningFigure}
\end{figure}

Next, Figure \ref{fig:LQRTuningFigure} shows the interception response with various values of the proportional gain $k_{\dot \psi}.$
Note that, as expected, the interception is successful for all values of the proportional gain $k_{\dot \psi}.$

\subsection{Monte Carlo analysis}
Next, we use the Monte Carlo analysis to compare the proposed range-based IOL guidance law, the line-of-sight-based IOL guidance law, and the standard proportional guidance (PG) law for point-mass dynamics.

The analysis has three parts: rear-aspect pursuer–evader scenarios, front-aspect scenarios, and interception against an evader executing evasive maneuvers.

In all scenarios, the evader has a constant thrust of 50 kN, a mass of 10,000 kg, and experiences gravity-induced normal acceleration. The pursuer is powered by a constant thrust of 15 kN. The PG law is implemented with a fixed gain of 3. An engagement is considered successful if the miss distance at intercept is less than 10 m \cite{aerospace10020155}.

To isolate guidance performance, sensor noise was not included. The pursuer’s initial speed exceeded the evader’s in all cases to model a terminal-phase intercept. The Monte Carlo setup follows the methodology described in \cite{PalumbHoming}.

\subsubsection{Rear Aspect Monte Carlo Analysis}
First, we evaluate the three guidance laws in a rear-aspect engagement geometry, where the pursuer is launched from behind the evader and moves in approximately the same direction, typically representing a tail-chase scenario.

We conduct 10,000 randomized Monte Carlo trials with initial conditions uniformly sampled from Table \ref{tab:MC_sim_rearapect_IC}.

\begin{table}[h!]
\centering
\begin{tabular}{|c|c|c|c|c|}
\hline
(min,max) & Velocity (m/s) & $\gamma$ (deg) & altitude (km) & downrange (km) \\
\hline
Pursuer & (800,1100) & (-45,45) & (12.5,20) & 0\\
Evader & (300,600) & (-45,45) & (10,20) & (5,10)\\
\hline
\end{tabular}
\caption{Range of uniformly distributed random initial conditions for Monte Carlo Simulation of rear aspect interception scenarios with point-mass dynamics.}
\label{tab:MC_sim_rearapect_IC}
\end{table}
Each table reports the distribution of interception times, miss distances, and closing velocities, along with the percentage of failed interceptions.
The results show that the line-of-sight-based IOL guidance achieves the lowest average miss distance (0.79 m) and an extremely low failure rate (0.04\%), demonstrating high robustness across varying initial geometries.
The range-based IOL guidance performs slightly worse, with a higher average miss distance (1.35 m) and a larger failure rate (1.52\%), suggesting that its effectiveness is more sensitive to initial condition variations.
The proportional guidance law shows performance similar to line-of-sight-based IOL, with an average miss distance of 0.81 m and a low failure rate (0.05 \%), though it exhibits slightly longer interception times on average.

Overall, these results indicate that while all three guidance laws achieve successful interceptions under most rear-aspect scenarios, the line-of-sight-based IOL law offers superior precision and reliability, outperforming both the range-based IOL and standard PG methods in terms of accuracy and robustness.

\begin{table*}[h!]
\centering
\resizebox{\textwidth}{!}
{%
\begin{tabular}{|c|ccc|ccc|ccc|}
\hline
 & \multicolumn{3}{c|}{\textbf{Line-of-sight-based IOL}} 
 & \multicolumn{3}{c|}{\textbf{Range based IOL}} 
 & \multicolumn{3}{c|}{\textbf{Proportional guidance}} \\
\hline
 & \makecell{Interception\\Time \\ (s)} & \makecell{Miss\\Distance \\ (m)} & \makecell{Closing\\Velocity \\ (m/s)} 
 & \makecell{Interception\\Time \\ (s)} & \makecell{Miss\\Distance \\ (m)} & \makecell{Closing\\Velocity \\ (m/s)} 
 & \makecell{Interception\\Time \\ (s)} & \makecell{Miss\\Distance \\ (m)} & \makecell{Closing\\Velocity \\ (m/s)} 
 \\
\hline
Average         & 11.03 & 0.79    & 952.8  & 11.07 & 1.35    & 917.6   & 11.18 & 0.81    & 977.8  \\
Median          & 10.87 & 0.58    & 965.1  & 10.91 & 0.57    & 926.7   & 11.00 & 0.60    & 978.7  \\
Variance        & 5.27  & 117.0   & 24864  & 5.27  & 155.2   & 20879   & 5.75  & 101.6   & 30880  \\
Minimum         & 5.62  & 4.15$\times 10^{-5}$ & 82.57  & 5.65  & 9.75$\times 10^{-5}$ & 193.1   & 5.65  & 2.09$\times 10^{-4}$ & 131.9  \\
Maximum         & 24.31 & 655.5   & 1908.1 & 24.08 & 655.5   & 1908.1  & 49.95 & 655.5   & 1818.1 \\
\hline
Percent Failure & \multicolumn{3}{c|}{0.04\%} 
                & \multicolumn{3}{c|}{1.52\%}
                & \multicolumn{3}{c|}{0.05\%} \\
\hline
\end{tabular}}
\caption{Comparison of Monte Carlo results for three guidance laws with point-mass dynamics under uniformly random rear aspect initial conditions from Table~\ref{tab:MC_sim_rearapect_IC}.}
\label{tab:combined_guidance_results_RA}
\end{table*}

\begin{figure}[h!]
    \centering
    \includegraphics[width=0.8\columnwidth]{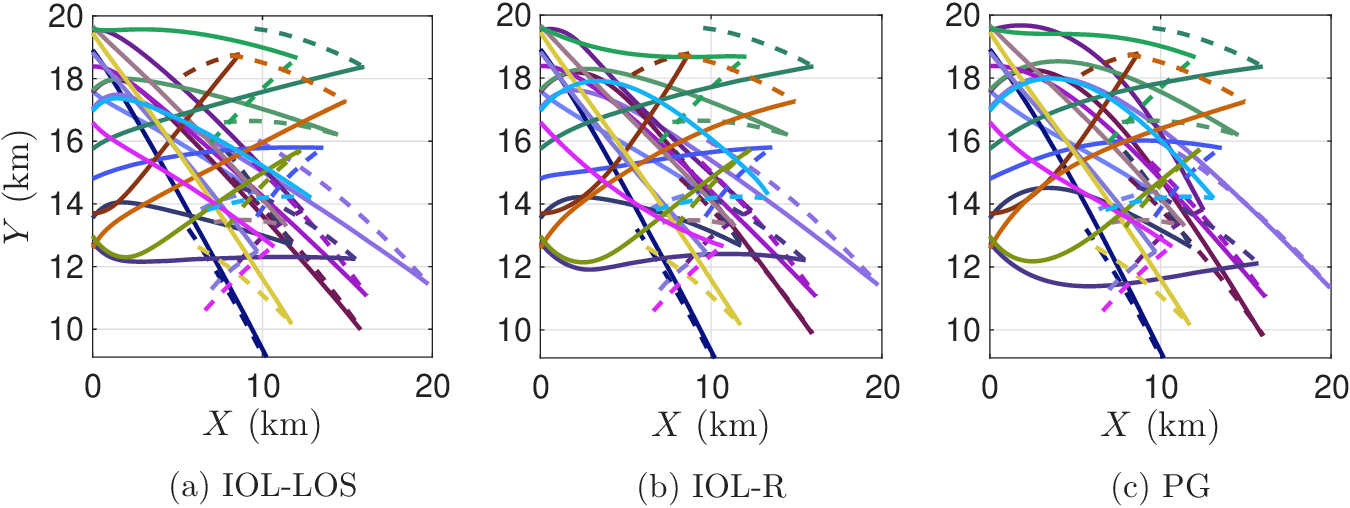}
    \caption{
    Pursuer (solid) and evader (dashed) trajectories in the rear aspect engagement scenario in 20 sample cases with the three guidance laws. }
    % 20 Samples of the Rear Aspect Monte Carlo Simulation for ((a) Line-of-sight-based IOL, (b) Range Based IOL, and (c) Proportional Guidance.  Dashed lines indicate the evader, and solid lines indicate the pursuer.}
    \label{fig:MC_PlotResultsRA}
\end{figure}

Figure \ref{fig:MC_PlotResultsRA} shows 20 sample trajectories from the rear-aspect Monte Carlo simulation for all three guidance strategies.
The initial conditions for each sample are identical across the guidance laws to enable a fair comparison.
Note that, among the strategies, the IOL-LOS-guided trajectories appear to result in the fastest intercepts.

\subsubsection{Front Aspect Monte Carlo Analysis}
Next, we evaluate the performance of the three guidance laws in a front-aspect engagement scenario. 
In this geometry, the pursuer missile is launched facing the evader’s direction of travel and initially moves in roughly the opposite direction of the target.
This setup represents a typical head-on engagement, where the missile experiences a high closing velocity but must execute precise maneuvers to intercept the target before the rapidly shrinking range allows it to strike.
To evaluate performance under realistic and varied conditions, we conduct a Monte Carlo simulation with 10,000 randomized trials.
Initial conditions are uniformly sampled from the ranges listed in Table \ref{tab:MC_sim_frontaspect_IC}.
\begin{table}[h!]
\centering
\small
\begin{tabular}{|c|c|c|c|c|}
\hline
(min,max) & Velocity ($\frac{\rm{m}}{\rm{s}}$) & $\gamma$ (deg) & altitude (km) & downrange (km) \\
\hline
Pursuer & (800,1100) & (120,240) & (10,30) & (15,20)\\
Evader & (300,600) & (-60,60) & (12.5,30) & 0\\
\hline
\end{tabular}
\caption{Range of uniformly distributed random initial conditions for Monte Carlo Simulation of front aspect interception scenarios with point-mass dynamics.}
\label{tab:MC_sim_frontaspect_IC}
\end{table}

\begin{figure}[h!]
    \centering
    \includegraphics[width=0.8\columnwidth]{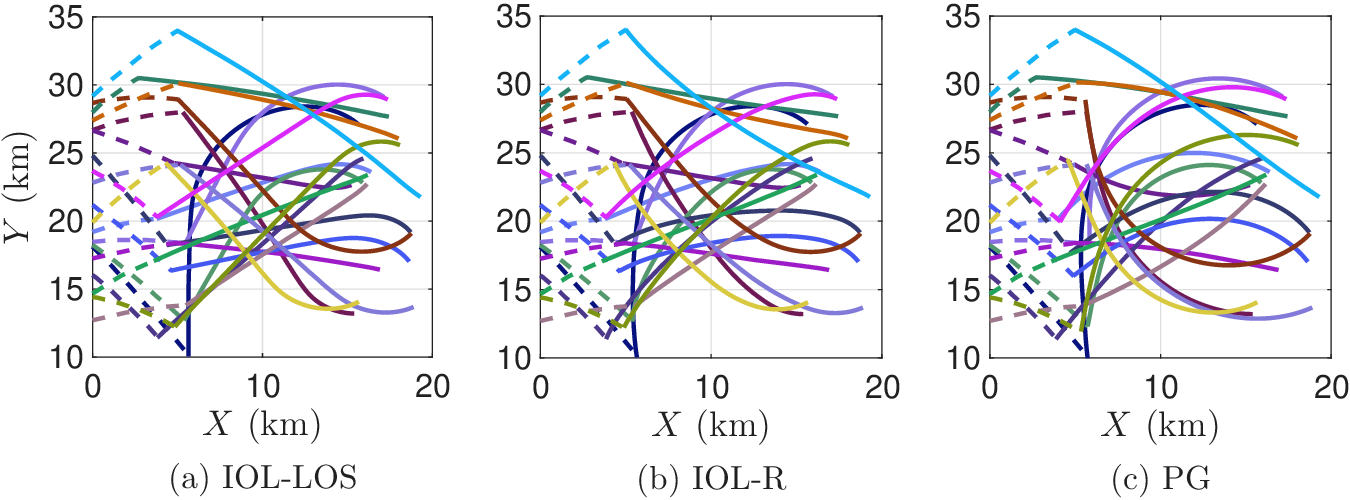}
    \caption{
    Pursuer (solid) and evader (dashed) trajectories in the front aspect engagement scenario in 20 sample cases with the three guidance laws.
    }
    % 20 Samples of the Front Aspect Monte Carlo Simulation for ((a) Line-of-sight-based IOL, (b) Range Based IOL, and (c) Proportional Guidance.  Dashed lines indicate the evader, and solid lines indicate the pursuer.}
    \label{fig:MC_PlotResultsFA}
\end{figure}

Table \ref{tab:combined_guidance_results_FA} summarizes the Monte Carlo results for the three guidance laws subject to inital conditions sampled from table \ref{tab:MC_sim_frontaspect_IC}. 20 samples from the Monte Carlo results are plotted in figure \ref{fig:MC_PlotResultsFA}.
The results show that the line-of-sight-based IOL guidance achieves the lowest average miss distance (0.79 m) and the lowest failure rate (0.15\%). 
The range-based IOL guidance performs worse, with a higher average miss distance (9.29 m) and a large failure rate (19.1\%).
The proportional guidance law shows worse performance to line-of-sight-based IOL, with an average miss distance of 8.44 m and a failure rate of (1.34 \%), it exhibits slightly longer interception times on average and larger variances across all metrics with respect to line-of-sight-based IOL.

Overall, these results indicate that the line-of-sight-based guidance and proportional guidance  were able to cause interceptions under most front-aspect scenarios; however, the line-of-sight-based IOL law offers superior precision and reliability, outperforming both the range-based IOL and standard PG method in terms of accuracy and robustness.

\begin{table*}[h!]
\centering
\resizebox{\textwidth}{!}
{%
\begin{tabular}{|c|ccc|ccc|ccc|}
\hline
 & \multicolumn{3}{c|}{\textbf{Line-of-sight-based IOL}} 
 & \multicolumn{3}{c|}{\textbf{Range based IOL}} 
 & \multicolumn{3}{c|}{\textbf{Proportional guidance}} \\
\hline
 & \makecell{Interception\\Time \\ (s)} & \makecell{Miss\\Distance \\ (m)} & \makecell{Closing\\Velocity \\ (m/s)} 
 & \makecell{Interception\\Time \\ (s)} & \makecell{Miss\\Distance \\ (m)} & \makecell{Closing\\Velocity \\ (m/s)} 
 & \makecell{Interception\\Time \\ (s)} & \makecell{Miss\\Distance \\ (m)} & \makecell{Closing\\Velocity \\ (m/s)} 
 \\
\hline
Average         & 12.33 & 2.75    & 1899  & 12.41 & 9.29    & 1846.8   & 12.84 & 8.44    & 1842.2  \\
Median          & 11.88 & 1.17    & 1931.1  & 12.09 & 1.62    & 1891.3   & 12.24 & 1.13    & 1878.4  \\
Variance        & 4.75  & 2666.5   & 46461  & 4.29  & 612.4   & 69425   & 7.31  & 8149.9   & 63266  \\
Minimum         & 8.13  & 3.86$\times 10^{-5}$ & 505.6  & 8.49  & 2.1$\times 10^{-4}$ & 566.1   & 8.15  & 4.9$\times 10^{-4}$ & 325.49  \\
Maximum         & 25 & 2457.1   & 2413.4 & 22.89 & 238.9   & 2357  & 42.42 & 3280.1   & 2466.7 \\
\hline
Percent Failure & \multicolumn{3}{c|}{0.15\%} 
                & \multicolumn{3}{c|}{19.1\%}
                & \multicolumn{3}{c|}{1.34\%} \\
\hline
\end{tabular}}
\caption{Comparison of Monte Carlo results for three guidance laws with point-mass dynamics under uniformly random front aspect initial conditions from Table~\ref{tab:MC_sim_frontaspect_IC}.}
\label{tab:combined_guidance_results_FA}
\end{table*}

Figure \ref{fig:MC_PlotResultsFA} shows 20 sample trajectories from the front-aspect Monte Carlo simulation for all three guidance strategies.
The initial conditions for each sample are identical across the guidance laws to enable a fair comparison.

\subsubsection{Front Aspect Manuevering Evader}
Next, we evaluate the performance of the three guidance laws in a front-aspect maneuvering evader engagement geometry.
This type of intercept problem typically represents a head-on engagement, where the missile benefits from a closing velocity advantage, but must maneuver precisely to intercept the rapidly decreasing relative range as the evader performs evasive maneuvers.
To assess the guidance laws under realistic and diverse conditions, we conduct a Monte Carlo simulation with 10,000 randomized trials. The inital conditions are sampled from table \ref{tab:MC_sim_frontaspect_Evasion_IC}. Where the column titled evasion time is the random time at which the evader begins performing an evasive pull of 10 g's upwards or downwards, chosen randomly.

\begin{table}[h!]
\centering
\small
\begin{tabular}{|c|c|c|c|c|c|}
\hline
(min,max) & Velocity ($\frac{\rm{m}}{\rm{s}}$) & $\gamma$ (deg) & altitude (km) & downrange (km) & Evasion Time (s)\\
\hline
Pursuer & (800,1100) & (157.5,202.5) & 10 & 10 & N/A \\
Evader & (300,600) & (-22.5,22.5) & 10 & 0 & (1,8)\\
\hline
\end{tabular}
\caption{Range of uniformly distributed random initial conditions for Monte Carlo Simulation of front aspect interception scenarios with point-mass dynamics.}
\label{tab:MC_sim_frontaspect_Evasion_IC}
\end{table}

\begin{figure}[h!]
    \centering
    \includegraphics[width=0.8\columnwidth]{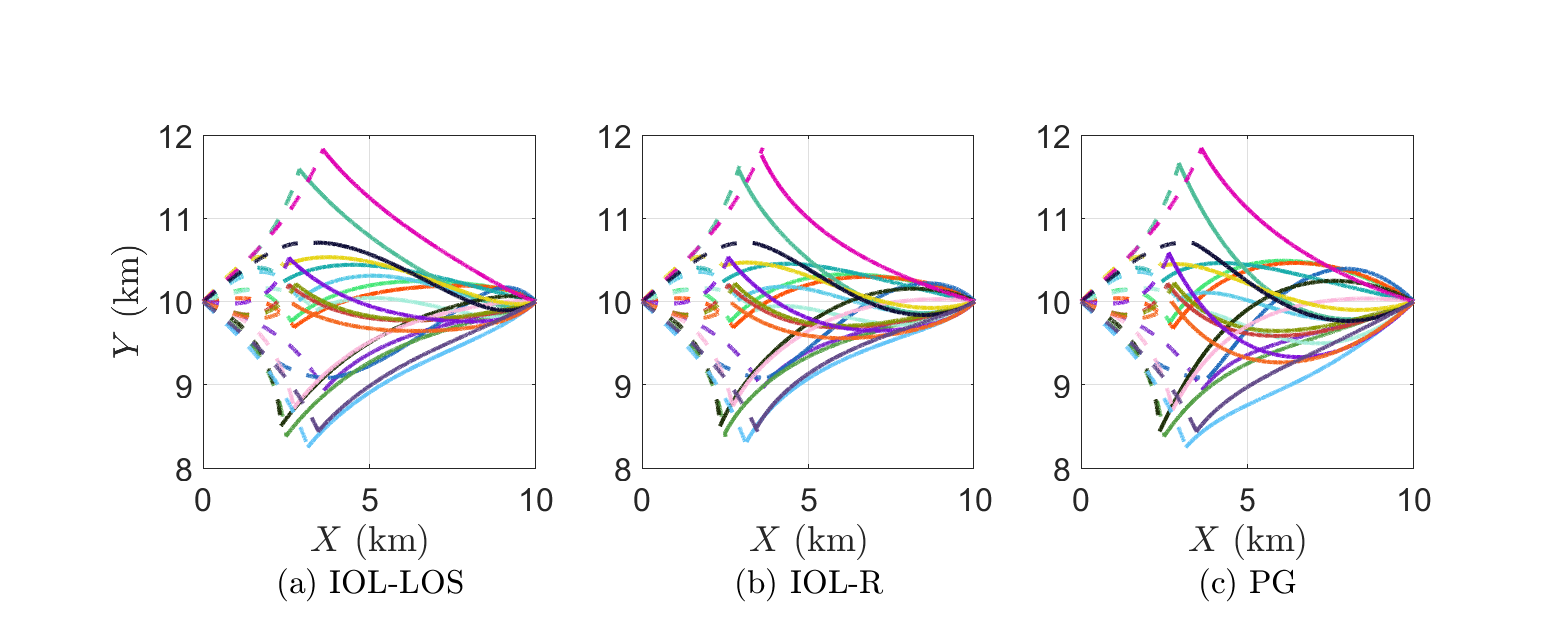}
    \caption
    {Pursuer (solid) and evader (dashed) trajectories in the front aspect engagement with the maneuvering evader scenario in 20 sample cases with the three guidance laws.}
    % {20 samples of the front aspect manuevering evader Monte Carlo Simulation for (a) Line-of-sight-based IOL, (b) Range Based IOL, and (c) Proportional Guidance.  Dashed lines indicate the evader, and solid lines indicate the pursuer.}
    \label{fig:MC_PlotResultsFA_Evasion}
\end{figure}

 The results are summarized in table \ref{tab:combined_guidance_results_FAE} and 20 samples are plotted in figure \ref{fig:MC_PlotResultsFA_Evasion}. The results show the line-of-sight-based IOL guidance achieves zero failures and suppior mean (0.927 m) and median (0.913 m) miss distances to the range-based IOL guidnace and proportional guidance. 
The range-based IOL guidance performed poorly with a failure rate of 42.75 \% indicating the sensitivity range-based IOL guidance has to inital geometry and evading manuervers.
The proportional guidance had a moderatly high percent failure (7.47\%) with a mean miss distance of 1.904 m and meadian miss distance of 1.261 m. The variance of all metrics are greater than the line-of-sight-based IOL.
Overall, The range-based IOL guidance is shown to not have acceptable performance. However, the strength of line-of-sight-based guidance in front aspect evading scenarios is demonstraighted with the zero failures compared proportional guidance's failure rate of 7.47\%. This is also seen in the lower interception time and miss distance metrics and the, on average, faster closing velocity that the line-of-sight-based IOL guidance maintains relative to proportional guidance.

\begin{table*}[h!]
\centering
\resizebox{\textwidth}{!}
{%
\begin{tabular}{|c|ccc|ccc|ccc|}
\hline
 & \multicolumn{3}{c|}{\textbf{Line-of-sight-based IOL}} 
 & \multicolumn{3}{c|}{\textbf{Range based IOL}} 
 & \multicolumn{3}{c|}{\textbf{Proportional guidance}} \\
\hline
 & \makecell{Interception\\Time \\ (s)} & \makecell{Miss\\Distance \\ (m)} & \makecell{Closing\\Velocity \\ (m/s)} 
 & \makecell{Interception\\Time \\ (s)} & \makecell{Miss\\Distance \\ (m)} & \makecell{Closing\\Velocity \\ (m/s)} 
 & \makecell{Interception\\Time \\ (s)} & \makecell{Miss\\Distance \\ (m)} & \makecell{Closing\\Velocity \\ (m/s)} 
 \\
\hline
Average         & 7.00 & 0.927    & 1394.5  & 7.05 & 14.62    & 1366.6   & 7.05 & 1.904    & 1365.3  \\
Median          & 6.95 & 0.913    & 1405.7  & 7.01 & 3.071    & 1382   & 7.00 & 1.261    & 1378.4  \\
Variance        & 0.30  & 0.243   & 25615  & 0.32  & 546.62   & 34080   & 0.32  & 5.216   & 32804  \\
Minimum         & 5.82  & 3.00$\times 10^{-5}$ & 898.87  & 5.89  & 0.0123 & 807.33   & 5.83  & 5.00$\times 10^{-4}$ & 799.33  \\
Maximum         & 8.88 & 2.119   & 1745.3 & 8.99 & 147.1   & 1739.6  & 9.16 & 25.88   & 1747.6 \\
\hline
Percent Failure & \multicolumn{3}{c|}{0.00\%} 
                & \multicolumn{3}{c|}{42.75\%}
                & \multicolumn{3}{c|}{7.47\%} \\
\hline
\end{tabular}}
\caption{Comparison of Monte Carlo results for three guidance laws with point-mass dynamics under uniformly random front aspect initial conditions with an evading target that randomly pulls 10 g's upwards or downwards at a random time from Table~\ref{tab:MC_sim_frontaspect_Evasion_IC}.}
\label{tab:combined_guidance_results_FAE}
\end{table*}
Figure \ref{fig:MC_PlotResultsFA_Evasion} shows 20 sample trajectories from the front-aspect maneuvering evader Monte Carlo simulation for all three guidance strategies.
The initial conditions for each sample are identical across the guidance laws to enable a fair comparison.

\begin{figure}[h!]
    \centering
    \includegraphics[width=0.8\columnwidth]{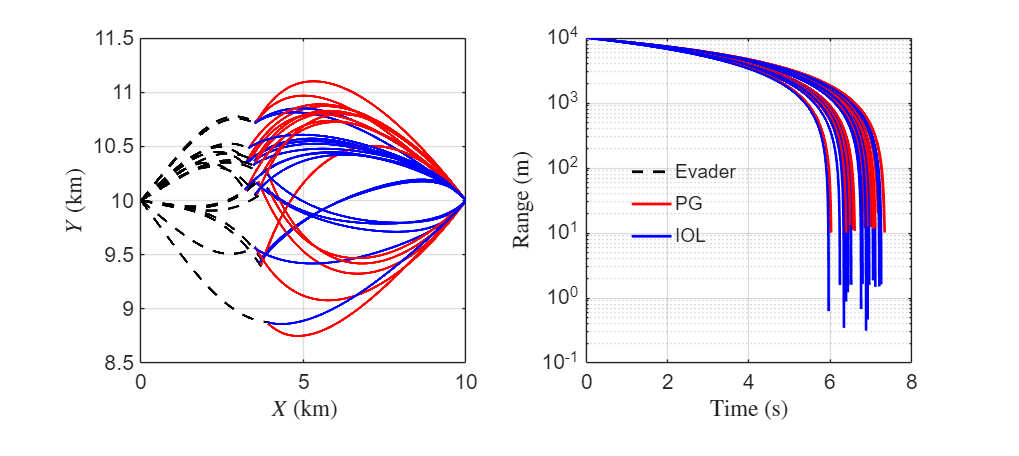}
    \caption{
    % Pursuer (solid) and evader (dashed) trajectories in the front aspect engagement with maneuvering evader scenario in 20 sample cases with the three guidance laws.}
    Results from front aspect maneuvering evader showing 10\% of proportional guidance's failures versus Line-of-sight-based IOL guidance with identical initial conditions, evasion time, and evasion direction.}
    \label{fig:PG_VS_IOL}
\end{figure}

Figure \ref{fig:PG_VS_IOL} shows the comparison of the behavior of proportional guidance and line-of-sight-based IOL guidance. 10\% of the failures of proportional guidance are sampled, and the two guidances are plotted against each other with identical initial conditions and evasion parameters for a fair comparison. The line-of-sight-based IOL guidance has faster interception times and a smaller miss distance when compared to proportional guidance. 

\section{Acknowledgment}
This research was supported by the Office of Naval Research grant N00014-23-1-2468.

\section{Conclusions}
This work presented two input–output feedback linearization (IOL)-based guidance laws for point-mass pursuer–evader interception scenarios using either range or line-of-sight (LOS) rate measurements. 
While both baseline approaches guarantee interception in ideal conditions, they exhibit practical implementation challenges—namely, singularities in the range-based formulation and divergence in the LOS-based formulation. To address these issues, a fuzzy blending strategy with proportional guidance was introduced for the range-based law, and a correction logic was proposed for the LOS-based law.

Monte Carlo simulations across rear-aspect, front-aspect, and front-aspect maneuvering target engagements demonstrated that the modified LOS-based IOL guidance law consistently achieved the highest accuracy, lowest miss distances, and minimal or zero failure rates, outperforming both the range-based IOL and standard proportional guidance laws. 
The range-based IOL guidance, while effective in many cases, proved more sensitive to initial conditions and target maneuvers, resulting in higher failure rates in challenging geometries.

Overall, the proposed modifications significantly improve the robustness and reliability of IOL-based guidance laws, particularly in dynamic and uncertain engagement scenarios.
Future work will focus on incorporating sensor noise models, actuator constraints, and three-dimensional engagement dynamics, as well as validating the proposed strategies in hardware-in-the-loop and flight test environments.

\appendix

\bibliography{Refs}

\end{document}